\documentstyle[prl,aps,amsfonts]{revtex}
\begin{document}
\draft
\twocolumn[%
\hsize\textwidth\columnwidth\hsize\csname@twocolumnfalse\endcsname
\title{Dynamical signatures of the vulcanization transition}
\author{Kurt Broderix\rlap,$^1$ 
        Paul M.~Goldbart\rlap,$^2$
        and 
        Annette Zippelius$^1$}
\address{$^1$Institut f\"ur Theoretische Physik, 
         Universit\"at G\"ottingen, 
         D-37073 G\"ottingen, Germany}
\address{$^2$Department of Physics, 
         University of Illinois at Urbana-Champaign, 
         1110 West Green Street, Urbana,
         Illinois 61801, USA}

\date{August 13, 1997}
\maketitle
\begin{abstract}
  Dynamical properties of vulcanized polymer networks are addressed
  via a Rouse-type model that incorporates the effect of permanent
  random crosslinks.  The incoherent intermediate scattering function
  is computed in the sol and gel phases, and at the vulcanization
  transition between them.  At any nonzero crosslink density within
  the sol phase Kohlrausch relaxation is found. The critical point is
  signalled by divergence of the longest time-scale, and at this point
  the scattering function decays algebraically, whereas within the gel
  phase it acquires a time-persistent part identified with the gel
  fraction.
\end{abstract}
\vspace*{-2.5pt}
]
Vulcanized matter has been suggested as a model substance for the
transition from a viscous fluid to an amorphous solid state, i.e.~the
glass transition.  Experimentally one has access to the static and
dynamic critical behavior, and theoretically a reliable
semi-microscopic model is available for actually computing properties
of the system in the amorphous state as well as close to the glass
transition. So far, the approach based on statistical mechanics has
mainly focused on static properties. Here we extend the analysis to
dynamic quantities and discuss the motion of a tagged monomer. We
thereby show that vulcanized systems display many of the {\it
dynamic\/} characteristics of {\it glassy\/} systems, in particular
a Kohlrausch law (generalized to nonzero wavenumbers) in the fluid
phase and a divergence of the longest time-scale as the glass
transition is approached.

The gelation transition has been discussed very successfully in terms
of a percolation transition~\cite{Reviews}.  The percolation threshold
is identified with the critical crosslink concentration, and the
average cluster size with the weight average molecular weight in the
sol phase.  The percolation picture has been confirmed by a
statistical-mechanical approach~\cite{Deam,Ball,Review}.  In addition,
the latter approach allows one to discuss structural and elastic
properties of the gel phase, which can not be obtained from a
percolation model unless additional assumptions are made.  The
shortcomings of percolation theory become even more apparent if one is
interested in dynamic quantities. There is obviously no dynamic
percolation model that could account for the observed anomalous
relaxation while approaching the gelation transition from the sol
phase.

Here we start from a (semi-)microscopic equation of motion, the Rouse
model~\cite{Doi}, and compute the incoherent intermediate scattering
function as the vulcanization transition is approached from the liquid
side.  We only consider permanently formed crosslinks and model them
by a harmonic potential.  We furthermore assume that each polymer is
equally likely to be crosslinked to each other polymer.  We do not
distinguish between gelation and vulcanization and use sol phase and
fluid phase as synonyms, because, as we shall see, the long time
dynamics is not affected by the molecular size and structure of the
building blocks of the random network.  Neglecting excluded volume
interactions, we analyse the dynamics by making use of the theory of
random graphs, created by Erd{\H o}s and R{\'e}nyi~\cite{Erdos1}. (For
an informal account of random graph theory see
Ref.~\cite{Erdos2}.)\thinspace\ Our main results are the following:

$\bullet$ We observe {\it three\/} distinct phases with 
  qualitatively different long-time dynamics.
  
$\bullet$ If the number of crosslinks $M$ is submacroscopic
  (i.e.~$o(N)$, where $N$ is the number of macromolecular chains in
  the uncrosslinked system) then the long-time behavior of the
  autocorrelation is diffusive: $S_{t}(\mbox{\boldmath$q$})\propto
  \exp\{-D_0\mbox{\boldmath$q$}^2t\}$.
  
$\bullet$ If the crosslink concentration $c:=M/N$ is non-zero but
  smaller than the critical value $c_{\text{crit}}$ for vulcanization
  then the time-decay of the autocorrelation is characterized by a
  spectrum of relaxation times, reflecting the diffusive motion of a
  distribution of cluster sizes.  The autocorrelation decays for large
  times as a {\it stretched\/} exponential:
  $S_{t}(\mbox{\boldmath$q$})\propto
  \exp\{-(t/\tau_{\mbox{\boldmath$\scriptstyle q$}})^{1/2}\}$ with,
  however, a scaling of time with wavenumber that is specific to a
  diffusion process: $\tau_{\mbox{\boldmath$\scriptstyle q$}}^{-1}
  \propto D_0 \mbox{\boldmath$q$}^2$. This relaxation is the true
  asymptotic behavior, and should not be confused with a formally
  similar time decay in the Rouse model for uncrosslinked systems,
  which occurs for time-scales smaller than the Rouse time~\cite{Doi}.
  
$\bullet$ As the vulcanization transition is approached, i.e.~for
  $c\uparrow c_{\text{crit}}$, the longest time-scale diverges like
  the inverse squared distance from the critical point, whereas the
  generalized diffusion constant goes to zero linearly and, more
  generally, is proportional to the weight average molecular weight.
  
$\bullet$ Precisely at $c_{\text{crit}}$ the distribution of
  relaxation times falls off algebraically for large relaxation times,
  implying a $(D_0\mbox{\boldmath$q$}^2t)^{-1/2}$ decay of the
  autocorrelation.
  
$\bullet$ In the gel phase, i.e.~for $c>c_{\text{crit}}$, the
  autocorrelation acquires a time-persistent part, which is identical
  to the gel fraction as obtained from static calculations.
  
$\bullet$ The mean square displacement grows linearly with time for
  all crosslink concentrations, with only a weak singularity at the
  vulcanization transition. Its long-time behavior is dominated by the
  \lq\lq mobile" monomers on finite clusters, even in the gel phase.
  
We consider a system of $N$ linear, identical, mono-disperse chains
of arc-length $L$ with force-free ends. Monomer $s$ on chain $i$ is
characterized by its time-dependent position vector
$\mbox{\boldmath$R$}_t(i,s)$ ($i=1,\ldots,N$ and $0\leq s \leq L$)
in $d$-dimensional space.  The simplest dynamics is purely
relaxational:
\begin{equation} \label{Eq1}
  \partial_t\mbox{\boldmath$R$}_t(i,s)
  = - \frac{1}{\zeta} \, 
      \frac{\delta H}{\delta \mbox{\boldmath$R$}_t(i,s)}
    + \mbox{\boldmath $\eta$}_t(i,s).
\end{equation}
The monomers' relaxation to the stationary state $\delta
H/\delta\mbox{\boldmath$R$}={\bf 0}$ with rate $\zeta$ is perturbed by
thermal fluctuations, which are modeled as Gaussian white noise
\mbox{\boldmath $\eta$} with zero mean and variance $ \left\langle
  \mbox{\boldmath$\eta$}^{\nu}_t(i,s)\, \mbox{\boldmath
    $\eta$}^{\mu}_{t'}(i',s') \right\rangle = {2}{\zeta}^{-1}\,
\delta^{\nu,\mu}\, \delta(t-t')\, \delta_{i,i'}\, \delta(s-s') $,
where $\nu,\mu=1,\dots,d$. We measure energies in units such that
$k_BT$ is unity. If the Hamiltonian were to consist only of the Wiener
term
\begin{equation} 
  H_W :=
  \frac{d}{2l}\sum_{i=1}^{N} \int_{0}^{L} \! {\rm d} s
  \left( 
    \frac{\partial\mbox{\boldmath$R$}_t(i,s)}{\partial s} 
  \right)^{2}
\end{equation}
then the dynamical equation would reduce to the Rouse model.  Here $l$
denotes the persistence length of the individual polymers.  In the
following we wish to treat a crosslinked melt or solution.  We only
consider the case of permanently formed crosslinks that constrain
randomly chosen pairs of monomers, and we study the relaxation of
monomers in the presence of a fixed crosslink configuration.  A
particular realization of $M$ crosslinks is characterized by a set of
indices $\{i_{e},s_{e};i'_{e},s'_{e} \}_{e=1}^{M}$ specifying which
monomer is connected to which other monomer.  We adopt here the
simplest distribution of crosslinks, i.e.~the crosslinks are
independent of each other and for each crosslink the pair $(i_e;i'_e)$
of polymer indices is realized with equal probability. We leave
unspecified the distribution of monomer labels $(s_e;s'_e)$ because
its form will not affect our results.  Thus, for instance, the case of
end-linking with unconstrained functionality is included.  The
crosslinks are modeled by a harmonic potential
\begin{equation} 
  U:=\frac{d}{2a^2}\,\sum_{e=1}^M 
  \left( 
    \mbox{\boldmath$R$}_t(i_e,s_e)-\mbox{\boldmath$R$}_t(i'_e,s'_e) 
  \right)^2,
\end{equation}
where $a>0$ controls the distance between monomers within a crosslink.
The Hamiltonian that we use is thus $ H := H_W + U$.  Hard
$\delta$-constraints for the crosslinks can be recovered from the
harmonic potential in the limit $a\to 0$~\cite{Solf95}.

The quantity of interest is the intermediate incoherent scattering
function
\begin{eqnarray} 
  S_t(\mbox{\boldmath$q$}) & := & 
  \lim_{t_o\to\infty}
  \frac{1}{N} \sum_{i=1}^N  
  \int_0^L \! \frac{{\rm d} s}{L} 
\nonumber \\ & &
  \big\langle \exp\!\left\{ 
    i\mbox{\boldmath$q$} \cdot 
    (\mbox{\boldmath$R$}_{t+t_o}(i,s)-\mbox{\boldmath$R$}_{t_o}(i,s))
   \right\}\big\rangle.
\end{eqnarray}
The above average refers to the noise $\mbox{\boldmath$\eta$}$ and is
taken with fixed given crosslink configuration.

As the Hamiltonian is quadratic the Langevin equation~(\ref{Eq1}) is
linear and $\mbox{\boldmath$R$}$ is a Gaussian Markov process, whose
distribution is in the limit $t_o\to\infty$ characterized by
\begin{eqnarray} 
& & 
  \frac{1}{2d}
  \left\langle
    \left( 
      \mbox{\boldmath$R$}_{t+t_o}(i,s) -
      \mbox{\boldmath$R$}_{t_o}(i,s) 
    \right)\left( 
      \mbox{\boldmath$R$}_{t+t_o}(i'\!,s') - 
      \mbox{\boldmath$R$}_{t_o}(i'\!,s') 
    \right)
  \right\rangle
\nonumber \\  &&\qquad=  
  \frac{1}{\zeta}\int_0^t \! {\rm d}\tau \,
  {\rm e}^{-\tau\Gamma}(i,s;i'\!,s') =: G_t(i,s;i',s'). 
\end{eqnarray}
Here ${\rm e}^{-\tau\Gamma}(i,s;i'\!,s')$ is the matrix representation
of the exponential of the linear operator $\Gamma$ which stems from
the deterministic part of Eq.~(\ref{Eq1}) and is defined by its
action, viz.,
\begin{eqnarray} 
  \lefteqn{ (\Gamma f)(i,s) := 
            -\frac{d}{\zeta l}\,
            \frac{\partial^2}{\partial s^2} f(i,s)
          }
\nonumber \\ & \displaystyle \label{Eq6}
  \hbox{} \qquad+ \frac{d}{\zeta a^2} \sum_{e=1}^M 
&
  \left(
    \delta_{i,i_e}\,\delta(s-s_e) - \delta_{i,i'_e}\,\delta(s-s'_e)
  \right)
\\ & & \hbox{} \qquad\quad\times
  \left(
    f(i_e,s_e) - f(i'_e,s'_e)
  \right),  
\nonumber
\end{eqnarray}
on functions $f$ over the configuration space obeying Neumann
boundary-conditions with respect to $s$.

We are only concerned with the long-time behavior of the scattering
function.  Thus we need the asymptotics of $G_t$ as $t\to\infty$,
which (because $\Gamma$ is non-negative definite by inspection) is
governed by the eigenspace of $\Gamma$ corresponding to zero
eigenvalues.  Moreover, $\Gamma f=0$ implies $(f,\Gamma f)=0$, and
this in turn implies (by virtue of Eq.~(\ref{Eq6}) and the boundary
conditions) that both $\partial f(i,s)/\partial s=0$ for all
$i=1,\dots,N$ and $f(i_e,s_e)=f(i'_e,s'_e)$ for all $e=1,\dots,M$. Let
us group the polymers into $K$ clusters $\{{\cal N}_k\}_{k=1}^K$,
which are defined as maximal path-wise connected components. Then,
according to the above reasoning, the eigenspace in question consists
of functions that are constant when restricted to any one cluster. We
therefore have, as $t\to\infty$,
\begin{equation} 
  G_t(i,s;i'\!,s') = D_0 t
  \sum_{k=1}^K \frac{1}{N_k}\, 
  \delta_{{\cal N}_k}(i)\, \delta_{{\cal N}_k}(i')
   + {\cal O}(1), 
\end{equation}
with the bare diffusion constant $D_0=1/(\zeta L)$. Here $N_k$ denotes
the number of polymers in the $k$-th cluster ${\cal N}_k$, whence
$\sum_{k=1}^KN_k=N$, and $\delta_{{\cal{}N}_k}(i)=1$ if $i\in{\cal
  N}_k$ and $\delta_{{\cal{}N}_k}(i)=0$ otherwise.

Due to the Gaussian nature of the process $\mbox{\boldmath$R$}$ we can
express the scattering function in terms of $G$, and find for the
long-time behavior of the scattering function
\begin{equation} \label{Eq8}  
  S_t(\mbox{\boldmath$q$})   
  = 
  \frac{1}{N} \sum_{k=1}^K N_k 
  \exp\!\left\{ 
     - \mbox{\boldmath$q$}^2\left( \frac{D_0 t}{N_k} 
     + {\cal O}_k(1)\right)
  \right\}.
\end{equation}
We have equipped the long-time corrections ${\cal O}_k(1)$ with the
cluster index $k$ to indicate that we are not allowed to handle them
as an overall pre-factor.

The above result~(\ref{Eq8}) is not at all surprising.  It merely
confirms the notion that the slowest modes correspond to the
center-of-mass diffusion of individual polymer clusters.  Here
$D_0/N_k$ may be interpreted as the diffusion constant of the $k$-th
cluster, and $N_k/N$ as the probability for a given site to belong to
the $k$-th cluster.  We remark that in accordance with this notion the
strength of the crosslink potential does not enter in Eq.~(\ref{Eq8}).
Thus the result covers the limiting case, $a\to 0$, of hard
crosslinks, too.

The statistics of the polymer clusters' sizes is determined by the
crosslink distribution.  Two different polymers are considered to be
linked if there is at least one crosslink joining any two monomers of
the respective polymers.  In our model we specify the total number of
crosslinks $M$ so that the total number of polymer-polymer links
varies in different crosslink realizations. These fluctuations are
suppressed~\cite{BGZ} in the macroscopic limit as there are ${\cal
  O}(N^2)$ possibilities for $M={\cal O}(N)$ crosslinks to form $M$
different polymer-polymer links and only ${\cal O}(N)$ other
possibilities.  For any given number of polymer-polymer links, the
statistics of the clusters' sizes can be taken over from
Ref.~\cite{Erdos1} by identifying polymer clusters with components of
random graphs.

As shown in Ref.~\cite{Erdos1}, the polymer clusters exhibit the
analogue of a percolation transition. Below the crosslink
concentration $c_{\text{crit}}=1/2$ there will be no clusters of
macroscopic size and almost all polymers are members of tree
clusters~\cite{Erdos3}, which are defined by the absence of loops.
For $c>\frac{1}{2}$ there will be ${\cal O}(N)$ polymers in exactly
one macroscopic cluster, which is not a tree, and almost all other
polymers still belong to tree clusters of submacroscopic
size~\cite{Erdos3a}.  Let $T_n$ denote the average number of
tree-clusters consisting of $n$ polymers.  Then, whatever the
crosslink concentration $c>0$ may be, one has~\cite{Erdos4}
\begin{equation} \label{Eq9}
  \lim_{N\to\infty}\frac{1}{N}\,T_n = \tau_n = 
\frac{1}{2c}\: \frac{n^{n-2}}{n!}
  \left(2c\,{\rm e}^{-2c}\right)^n.
\end{equation}
In the macroscopic limit, the average fraction ${\cal S}_\infty$ of
polymers in the giant cluster is given by~\cite{Erdos5}
\begin{equation} 
   {\cal S}_\infty =
   1 - \sum_{n=1}^\infty n\,\tau_n
\end{equation}
which, for $c\le\frac{1}{2}$, amounts to 
${\cal S}_\infty=0$~\cite{Erdos6}.

To work out the long-time behavior of the autocorrelation, given the
statistics of polymer clusters' sizes, we shall first make the
simplifying assumption that we can ignore the possible dependence of
the long-time correction ${\cal O}_k(1)$ on the index $k$ of the
polymer cluster.  This assumption can indeed be justified~\cite{BGZ}
for finite cluster sizes.  From Eqs.~(\ref{Eq8}) and (\ref{Eq9}) we
get for the scattering function, averaged over the distribution of
crosslinks
\begin{equation} \label{Eq11}
  S_t(\mbox{\boldmath$q$})   
  =
  {\cal S}_\infty
  + \sum_{n=1}^\infty n\,\tau_n
  \exp\!\left\{ 
     -D_0 \mbox{\boldmath$q$}^2 t/n
  \right\}.
\end{equation}
This result holds for all $c>0$ with ${\cal S}_\infty=0$ for $0\le
c\le \frac{1}{2}$. In fact, one would expect
$S_t(\mbox{\boldmath$q$})$ to be self-averaging~\cite{BGZ}.

We now discuss four distinct regimes:

$\bullet$ For a {\it submacroscopic\/} number of crosslinks there is
  only a submacroscopic number of clusters, which do not contribute in
  the thermodynamic limit. Hence the scattering function decays, to
  leading order, according to the Rouse model:
  $S_t(\mbox{\boldmath$q$})=\exp\{-D_0\mbox{\boldmath$q$}^2t\}$.
  
$\bullet$ In the sol phase with a {\it non-zero\/} concentration of
  crosslinks, i.e.~for $0<c<\frac{1}{2}$, the distribution of cluster
  sizes falls off exponentially for large $n$:
  \begin{equation} \label{Eq12}
    n\tau_n\sim\frac{1}{2c\sqrt{2\pi}}\,n^{-3/2}\,\exp\{-nh(c)\}, 
  \end{equation}
  with $h(c):=2c-1-\ln(2c)>0$.  This exponential distribution of
  cluster sizes---and hence diffusive time-scales---gives rise to a
  stretched exponential decay of the scattering function in the sol
  phase:
  \begin{equation} \label{Eq13}
    S_t(\mbox{\boldmath$q$}) \sim
    \frac{1}{(8c^2D_0\mbox{\boldmath$q$}^2t)^{1/2}}
    \exp\{-2(h(c)D_0\mbox{\boldmath$q$}^2t)^{1/2}\}
  \end{equation}
  for $D_0\mbox{\boldmath$q$}^2t\to\infty$.  The time-scale of the
  stretched exponential diverges like
  $\left(c-\frac{1}{2}\right)^{-2}$ as the vulcanization transition is
  approached, whereas the generalized diffusion constant,
  \begin{equation}
    D_{\text{eff}}^{-1} := 
    \lim_{\mbox{\boldmath$\scriptstyle q$} \to 0} \mbox{\boldmath$q$}^2 
    \int_0^{\infty}\!{\rm d}t\,
    S_t(\mbox{\boldmath$q$}) =
    \frac{D_0^{-1}}{1-2c}, 
  \end{equation}
  goes to zero linearly as $c\uparrow\frac{1}{2}$.
  
$\bullet$ At the {\it critical point\/} for vulcanization, i.e.~at
  $c=\frac{1}{2}$, the distribution of cluster sizes falls off
  algebraically, $n \tau_n\propto n^{-3/2}$. Consequently the
  autocorrelation shows algebraic decay in time:
  $S_t(\mbox{\boldmath$q$})\propto(D_0\mbox{\boldmath$q$}^2t)^{-1/2}$.
  
$\bullet$ In the {\it gel phase\/}, i.e.~for $c>\frac{1}{2}$, the
  scattering function acquires a time-persistent part ${\cal
    S}_\infty$, which is given~\cite{Erdos6} by the largest root of
  $1-{\cal S}_\infty =\exp\big(-2c\,{\cal S}_\infty\big)$.  As
  expected, the time-persistent part of the autocorrelation is just
  the fraction of polymers in the macroscopic cluster, i.e.~the gel
  fraction.  The distribution (\ref{Eq9}) of finite cluster sizes is
  valid on both sides of the transition~\cite{Stauffer}. Hence we
  expect to see the stretched exponential behaviour (\ref{Eq13}) for
  the decaying part of the scattering function in the gel phase as
  well.
    
Lastly, we mention that the time-delayed square monomer displacement,
\begin{equation}
  C_t :=\frac{1}{N}\sum_{i=1}^N \int_0^L\!{\rm d}s\, 
  G_t(i,s;i,s) \sim 2Dt, 
\end{equation} 
increases asymptotically linearly with time for all crosslink
concentrations $c$. In terms of polymer clusters the proportionality
constant $D$ is given by $D=D_0 K/N=D_0 \sum_{n=1}^{\infty}\tau_n$.
This changes smoothly with crosslink concentration: in the sol phase
it is given by $D=D_0 (1-c)$; at $c=1/2$ it is twice continuously
differentiable; deep in the gel phase we find $D\sim D_0\exp(-2c)$.
Hence the squared monomer displacement is unsuitable for detecting the
sol-gel transition because its long-time behavior is always dominated
by the finite polymer clusters. 

We conclude with a few remarks concerning the experimental situation.
Relaxation following a stretched exponential
$\exp\{-(t/\tau)^{\beta}\}$ seems to be a {\it universal\/} feature of
glassy systems. It has been observed in fluids undergoing either
thermal or chemical vitrification, as well as in magnetic systems
undergoing a spin-glass transition.  It can always be interpreted in
terms of a distribution of time-scales, although the nature of the
underlying elementary process may be quite different.  In the present
model the elementary dynamical process is diffusion, and the stretched
exponential results from a superposition of many diffusive modes,
reflecting the distribution of cluster sizes~\cite{Polydisp}.  This
gives rise to a highly characteristic $\mbox{\boldmath$q$}$-dependence
of the time-scale of the stretched exponential, namely
$\tau_{\mbox{\boldmath$\scriptstyle q$}}^{-1} \propto
D_0{\mbox{\boldmath$q$}}^2$. Such a time-wavenumber scaling has indeed
been seen in neutron spin-echo experiments on inorganic and organic
glasses~\cite{MezeiRichter}.

As the gelation transition is approached one observes experimentally a
divergence of the longest time-scale, and at the critical point
density fluctuations are found to decay algebraically.  We compare our
results to the data of Martin et al.~\cite{Martin91}. These authors
use quasi-elastic light scattering to study silica sol-gels in the
pre- and post-gel regimes. They observe a stretched exponential in the
sol phase with an exponent $\beta \approx 0.66$ and, furthermore,
relate the time-scale $\tau$ of the stretched exponential to a
diffusive time scale.  The latter is found to diverge as the gel point
is approached with an exponent close to $2.2$.  In the critical gel
they find algebraic decay of density fluctuations with an exponent
value of $\approx 0.35$.  All these findings agree qualitatively with
the predictions of our model. In fact, our expression~(\ref{Eq8}) for
$S_t(\mbox{\boldmath$q$})$ has been suggested on phenomenological
grounds as a starting point for a discussion of the critical dynamics
at the sol-gel transition \cite{Pheno}. Although a general scaling of
the diffusion constant of a individual cluster with its size was
postulated, it has turned out that the so-called Rouse limit, which
leads back to our representation (\ref{Eq8}), agrees well with
experiment~\cite{Martin91}.  In comparing our results with experiment
one should keep in mind that we have used a distribution of crosslinks
lacking any correlations, and have not taken into account excluded
volume interactions.  Presumably this amounts to a mean-field
approximation of a short-range model.  Vulcanization of a dense melt
of linear macromolecules has been argued~\cite{deGennes77} to be
mean-field--like, except within a narrow critical region of crosslink
densities whose width decreases like the one third power of the ratio
of the persistence length to the chain length. Since the argument
includes and is in fact based on the distribution of cluster sizes, it
should also hold for the dynamics.

\smallskip
\noindent{\it Acknowledgments\/} -- We thank R.\ Kree for useful 
discussions. This work was supported by U.S.~National Science
Foundation Grant DMR94-24511 (PG), by NATO CRG~94090 (AZ,PG), and by
the DFG through SFB 345 (AZ).


\begin{references}
\bibitem{Polydisp} More generally, only the total length of all chains
  linked to a cluster determines the asymptotic dynamics. Hence our
  conclusions are not restricted to vulcanized chains but, e.g., hold
  equally well for polydisperse melts with a length distribution
  falling exponentially.
\end{references}
\end{document}